\documentclass[journal,10pt,twocolumn]{IEEEtran}
\pdfoutput=1 
\usepackage{setspace}
\usepackage{amssymb,amsmath,latexsym}
\usepackage{graphicx,subfigure}
\usepackage{color,cite,times}
\usepackage{epstopdf}
\usepackage{multirow}
\usepackage{array}

\usepackage{amsmath}
\usepackage{bm}
\usepackage{algorithm}
\usepackage{algorithmic}
\usepackage[T1]{fontenc}

\usepackage{booktabs}
\usepackage{bbm} 
\usepackage{subfigure}

\begin{document}
	\title{Energy-Efficient UAV-Mounted RIS Assisted Mobile Edge Computing}
	\author{Zhiyuan Zhai, Xinhong Dai, Bin Duo,~\IEEEmembership{Member,~IEEE}, Xin Wang,~\IEEEmembership{Senior~Member,~IEEE} and Xiaojun~Yuan,~\IEEEmembership{Senior~Member,~IEEE} 
			
		\thanks{Z.~Zhai, X.~Dai B.~Duo and X.~Yuan are with the the University of Electronic Science and Technology of China, Chengdu 611731, China. X.~Wang is with the  Fudan University, Shanghai 200433, China.
			

		}
	}
	\maketitle
	
	\begin{abstract}
		Unmanned aerial vehicle (UAV) and reconfigurable intelligent surface (RIS) have been recently applied in the field of mobile edge computing (MEC) to improve the data exchange environment by proactively changing the wireless channels through maneuverable location deployment and intelligent signals reflection, respectively. Nevertheless, they may suffer from inherent limitations in practical scenarios. UAV-mounted RIS (U-RIS), as a promising integrated approach, can combine the advantages of UAV and RIS to break the limit. Inspired by this, we consider a novel U-RIS assisted MEC system, where a U-RIS is deployed to assist the communication between the ground users and an MEC server. The joint UAV trajectory, RIS passive beamforming and MEC resource allocation design is developed to maximize the energy efficiency (EE) of the system. To tackle the intractable non-convex problem, we divide it into two subproblems and solve them iteratively based on successive convex approximation (SCA) and the Dinkelbach method. Finally we obtain a high-performance suboptimal solution. Simulation results show that the proposed algorithm significantly improves the energy efficiency of the MEC system.
		
	\end{abstract}
	\section{Introduction}
	Driven by the popularity of mobile users and unprecedented increase of network traffic in the Internet of Things (IoT)\cite{palattella2016internet}, mobile edge computing (MEC) is regarded as an emerging paradigm that executes computation-intensive and latency-critical tasks at the network edges to meet the demands of the resource-constrained mobile devices\cite{mao2017survey}. However, imperfect offloading links limit the exploitation of MEC. For example, since the direct offloading link may be blocked with a high probability, the poor channel condition forces the users to  process their tasks locally to maintain strict lantency requirements, which is unbearable for resource-limited mobile users. Hence, many works aim to improve the channel quality of  MEC systems.
	Owing to the line-of-sight (LoS) transmission and flexible maneuverability, it is promising to enable unmanned aerial vehicles (UAVs) to carry an MEC server (UAV-carried server) for providing data exchange and computing services\cite{li2020energy}, \cite{xu2021uav}. Particularly,\cite{li2020energy} proposed a joint resource allocation and UAV trajectory optimization scheme to maximize the  energy efficiency (EE), and \cite{xu2021uav} deployed multiple ground servers and a UAV server in a cooperative manner to provide high-quality edge computing. As another way to improve the channel quality,
	reconfigurable intelligent surface (RIS) is a cost-effective and energy-efficient equipment 
	that can be manipulated to alter the incident signal.
	It is considered as a win-win strategy to integrate RIS into MEC systems\cite{mukherjee2021interplay}. The work in\cite{bai2020latency} reported
	that up to 20$\%$ of the computing latency can be eliminated with the existence of RIS.
	
	Unfortunately, both UAV and RIS face their respective deficiencies, e.g., finite endurance and limited coverage. To overcome these issues, mounting RIS on UAV (named UAV-mounted RIS, U-RIS) to support terrestrial communication\cite{you2021enabling} is a promising approach.
	Compared with the UAV-carried server scheme, the U-RIS structure can be regarded as an effective upgrade to the traditional land server-based MEC system, without the need for routing change and system reconstruction. Furthermore, a
	 RIS is usually much lighter than an MEC server\cite{dai2020reconfigurable},
	leading to a lower UAV's on-board energy consumption. Besides, in the RIS-assisted MEC scheme, the RIS coated on the facade of buildings is only effective for the users in the front space. By contrast, on the aerial platform of UAV, RIS can enjoy a better full-angle panoramic beamforming capability towards users.
	
	For the above reasons, we consider a U-RIS assisted MEC system. Due to complex practical environments,
	the link between the ground users and the MEC server may be blocked. A U-RIS is dispatched to assist the users in offloading their tasks where the signals of users are reflected to the ground MEC server via the U-RIS.
	To balance the total processing bits and the energy consumption, our goal is to maximize the energy efficiency of the propposed system by jointly optimizing the UAV trajectory, passive beamforming, and resource allocation. 
	The formulated problem is a mixed-integer non-linear fractional programming problem. By leveraging the successive convex approximation (SCA) technique and the Dinkelbach method, we develop an efficient algorithm to obtain a suboptimal solution. 
	Numerical results demonstrate that the proposed algorithm achieves a substantial EE improvement of the MEC systems over other baseline schemes, and significant trade-off is made in the flight trajectory optimization.
	\section{System Model and Problem Formulation}
	
	\begin{figure}[tp]
		\setlength{\abovecaptionskip}{0.1cm} 
		\centering
		\includegraphics[width=2.6in,height=2.8cm]{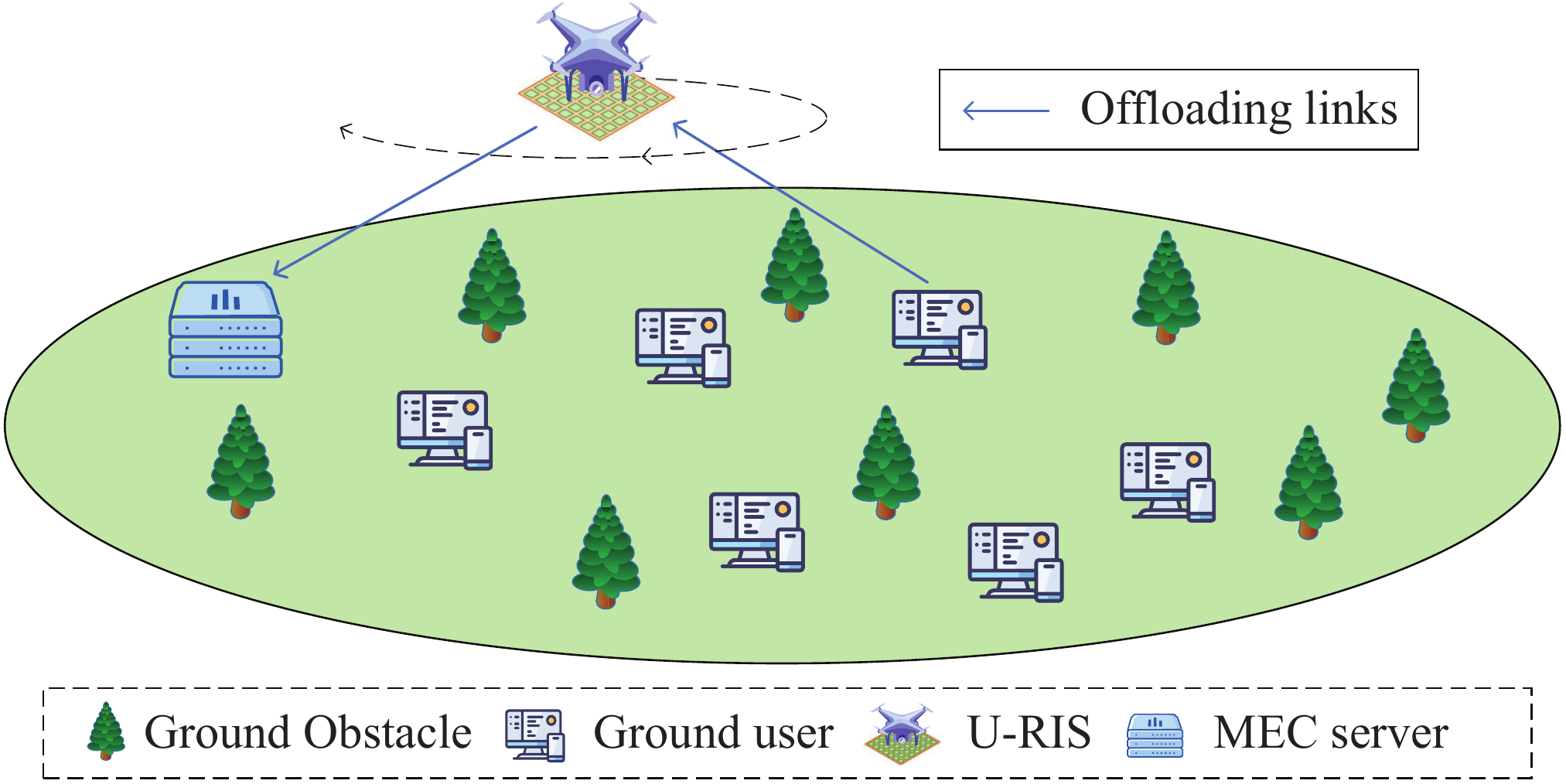}
		\caption{A U-RIS assisted MEC system.}
		\label{figure_1}
		\vspace{-4mm}
	\end{figure}

	The U-RIS assisted MEC network is shown in Fig.~\ref{figure_1}, consisting of $K$ users, an MEC server and a RIS that is mounted on a UAV. The U-RIS is deployed to assist the MEC server in providing edge computing service to the ground users.
	Denote by $\mathcal{K}\triangleq\{1,2\cdots,K\}$ the set of users. Without loss of generality, the time span $\emph{T}$ is divided into $\emph{N}$ time slots with size of ${\delta_t}$, which is indexed by $\mathcal{N}\triangleq\{1,2\cdots,N\}$. 
	
	It is assumed that all the nodes in the MEC system are located in the three-dimensional (3D) Cartesian coordinate system, and the horizontal coordinates of the MEC server as well as user $\emph{k}$ can be denoted by ${\bf w}_{\text s}=[x_{\text s},y_{\text s}]$ and ${\bf w}_k=[x_k,y_k]$, respectively. We assume that the U-RIS flies at a fixed altitude $\emph{H}$ and its position remains unchanged within each time slot. Thus, the horizontal trajectory of the U-RIS during the time span $\emph{T}$ can be denoted by the sequence ${\bf q}[n]=[x[n],y[n]], n \in \mathcal{N}$, satisfying the following mechanical and flight constraints
	\begin{subequations} \label{mobility.1}
		\begin{align}
			& {\bf v}[n]=\frac{{\bf q}[n+1]-{\bf q}[n]}{\delta_t},\left|\left|{\bf v}[n]\right|\right|\leq { v}_{max},\forall n,\label{mobility.1.a}\\[-0.5mm]   	
			& \boldsymbol {a}[n]=\frac{{\bf v}[n+1]-{\bf v}[n]}{\delta_t},\left|\left|{\boldsymbol{a}}[n]\right|\right|\leq {a}_{max},\forall n, \label{mobility.1.b}\\
			&{\bf q}[1]={\bf q}_0,{\bf q}[N+1]={\bf q}_F,\left|\left|{\bf q}[n]\right|\right|\leq r_d,\forall n,\label{mobility.1.e}
		\end{align}
	\end{subequations}
	where ${ v}_{max}$ and ${{a}}_{max}$ are the UAV's maximum speed and acceleration, ${\bf q}_0$ and ${\bf q}_F$ denote the initial and final horizontal positions of the U-RIS, and $r_d$ is the horizontal flight range of the U-RIS. The last part of (\ref{mobility.1.e}) is owing to that the U-RIS should be practically tethered to a fixed ground station for dependable power supply and  stable control\cite{you2021enabling}. 
	
	\subsection{Communication Model}
	
	We assume that the MEC server and the ground users are equipped with a single omni-directional antenna for each. The U-RIS is equipped with $M=M_x \times M_y$ reflective elements, forming an $M_x \times M_y$ uniform rectangular array (URA). 
	Let $\theta_i[n] \in \left[0,2\pi\right), i \in \mathcal{M}\triangleq\{1,\cdots, M\}$ denote the phase of the $i$th reflecting element in time slot $n$, and $\boldsymbol{\Theta}[n]=diag\{ e^{j\theta_1[n]},  e^{j\theta_2[n]},\cdots, e^{j\theta_M[n]}\}$ be the diagonal phase array for the RIS in the $n$th time slot.
	
	 We assume the links from the ground users to the U-RIS (G-U link) and from the U-RIS to the MEC server (U-S link) follow a quasi-static block fading LoS model\cite{zeng2017energy}. The link between user ${k}$ and the U-RIS in the $n$th time slot, denoted by ${\bf h}_{k}[n] \in \mathbb{C}^{M\times 1}$, can be expressed as\cite{li2021robust}
	\begin{equation}\label{channel.1}
			\setlength{\abovedisplayskip}{2pt}
		\setlength{\belowdisplayskip}{1pt}
		{\bf h}_{k}[n]=\tau[n]\boldsymbol{a}^k_x[n]\otimes\boldsymbol{a}_y^k[n],
		\setlength{\belowdisplayskip}{2pt}
	\end{equation}
	where
	\begin{align*}
		\boldsymbol{a}^k_x[n]&=[1,e^{-j\frac{2\pi}{\lambda}d\cos\phi_{k}[n]\sin\varphi_{k}[n]},\ldots,\\[-1mm]
		&\quad\quad\quad  e^{-j\frac{2\pi}{\lambda}(M_x-1)d\cos\phi_{k}[n]\sin\varphi_{k}[n]}]^T,\\[-1mm]
		\boldsymbol{a}^k_y[n]&=[1,e^{-j\frac{2\pi}{\lambda}d\sin\phi_{k}[n]\sin\varphi_{k}[n]},\ldots,\\[-1mm]
		&\quad\quad\quad e^{-j\frac{2\pi}{\lambda}(M_y-1)d\sin\phi_{k}[n]\sin\varphi_{k}[n]}]^T,\\[-1mm]
		&\sin\phi_{k}[n]\sin\varphi_{k}[n]=\frac{y[n]-y_k}{d_k[n]},\\[-0.8mm]
		&\cos\phi_{k}[n]\sin\varphi_{k}[n]=\frac{x[n]-x_k}{d_k[n]},
	\end{align*}
	$\tau[n]=\sqrt{\beta_{0}d_{k}^{-\alpha_{L}}[n]}$, where $\beta_{0}$ is the channel gain with a distance of 1 meter, $d_k[n]=\sqrt{\|({\bf q}[n]-{\bf w}_k)\|^2+H^2}$ is the distance between user $k$ and the U-RIS in time slot $n$, $\alpha_L$ is the pass loss exponent due to the LoS transmission. Moreover, $\phi_{k}[n]$ and $\varphi_{k}[n]$ denote respectively the azimuth and elevation angles of user $k$ at time slot $n$, $\lambda$ is the wavelength of carrier and $d$ is the antenna separation. The U-S link in time slot $n$, denoted as ${\bf h}_{{\text s}}[n]$, can be modeled similarly.
	
	We consider the TDMA scheme for the task offloading, which means that only one user communicates with the server in a time slot. Denote $c_k[n]$ as the user scheduling variable. If user $k$ is chosen to be served by the U-RIS in the $n$th time slot, we have $c_k[n]$=1, yielding the following constraints
	\begin{equation}
		\sum_{k=1}^K c_k[n]=1,\,
		c_k[n] \in \{0,1\},\forall k,n.\label{sel.2}
	\end{equation}
	Therefore, the achievable rate of user $k$ in the $n$th time slot can be expressed as
	\begin{equation}\label{rate.1}
		\setlength{\abovedisplayskip}{1pt}
		R_k[n]=c_k[n]B\log_2(1+\frac{P_k{\left|({\bf h}_{\text s}[n])^{H}{\boldsymbol \Theta}[n]{\bf h}_k[n]\right|}^2}{\sigma^2}),
	\end{equation}
	where the parameter $B$, $P_k$ and ${\sigma^2}$ denote the channel bandwidth, fixed transmission power of users and the noise variance respectively. Thus, the average achievable rate of user $k$ during the computing cycle $T$ is given by $R_k=\frac{1}{N}\begin{matrix}
		\sum_{n=1}^NR_k[n]
	\end{matrix}$.
	\subsection{Computation Model}
	To exploit the full granularity in task partitioning and computing resources, we consider the way of partial offloading\cite{jeong2017mobile}. Specifically, the computing tasks can be divided into arbitrary sizes, and part of the tasks can be offloaded to the server, while the remaining tasks are processed locally. Denote the total offloading and local computing tasks of user $k$ over $N$ time slots as ${l^{\text{o}}_{k}}$ and ${l^{\text{l}}_{k}}$ in bits, respectively. 
	Let $T_k$ be the maximum tolerable latency of user $k$. Then we have
	\begin{align}\label{toler.1}
		T_k\geq \max\left\{\frac{{l^{\text{l}}_{k}}\chi_k}{f^{\text{l}}_k},\frac{{l^{\text{o}}_{k}}\chi_k}{f^{\text{o}}_k}+\frac{{l^{\text{o}}_{k}}}{R_k}\right\},\forall k,
	\end{align}
	where $\chi_k$ denotes the number of CPU cycles required for processing one bit of user $k$, $f^{\text{l}}_k$ is user $k$'s fixed CPU frequency,  and $f^{\text{o}}_k$ is the allocated CPU frequency to compute the task of user $k$ at the MEC server. This constraint is based on two assumptions: First, the edge computing for user $k$ does not start until $l_k^o$ bits are offloaded; second, using  dynamic
	voltage and frequency scaling
	(DVFS) technique, the server can dynamically allocate its  resources. 
	\subsection{Energy Consumption Model}
	In the MEC system, the total energy consumption over $N$ time slots is composed of three main parts: the energy consumed by the users for offloading and local computing,  by the server for computing, and by the U-RIS flight. The energy consumption for user $k$ can be formulated as\footnote{Actually, the offloading duration is ${{l^{\text{o}}_{k}}}/{R_k}$, but $T_k$  is mainly occupied by ${{l^{\text{o}}_{k}}}/{R_k}$  and the energy used for transmission is relatively small\cite{li2020energy}. So we can replace it with $T_k$ to simplify the model without performance loss.}
	\begin{equation}\label{energy.1}
		\setlength{\belowdisplayskip}{3pt}
		E_{k}^{\text{u}}={T_k}P_k+\varphi_{\text{u}} \chi_k{l^{\text{l}}_{k}}(f^{\text{l}}_k)^{2},\forall k,
	\end{equation}
	where ${\varphi_{\text{u}}}$ is the switched capacitance coefficient for the users \cite{mei2019joint}. Denote ${\varphi_{\text{s}}}$ as the coefficient for the server, and the energy consumption of the server for computing user $k$'s tasks is 
	\begin{align}\label{energy.2}
		E_k^{\text{s}}={\varphi_{\text{s}}}\chi_k {l^{\text{o}}_{k}}(f^{\text{o}}_k)^{2},\forall k.
	\end{align}
	
	We adopt the novel energy consumption model for rotary-wing UAVs proposed in \cite{2202.08486}, which takes the practical thrust-to-weight ratio into consideration, i.e.,
	\begin{align}\label{energy.3}
		&\mathit{E_{\text{p}}}[n] 
		=\mathit{P}_{0}\Big(1+\frac{3||\mathbf{v}[n]||^2}{U_{tip}^{2} }\Big) +\frac{1}{2}d_{0}\rho sA||\mathbf{v}[n]||^3\notag \\&+ \mathit{P_{i}}\kappa[n]\sqrt{((\kappa[n])^2+\frac{||\mathbf{v}[n]||^4}{4v_{0}^{4} } )^{\frac{1}{2} } -\frac{||\mathbf{v}[n]||^2}{2v_{0}^{2}}},\forall n,    
	\end{align}
	where $\mathit{E_{\text{p}}}[n]$ is a factor of the energy consumption for the U-RIS flight during the $n$th time slot, and $\kappa[n]=({1+\frac{4m||\mathbf{a}[n]||^2+\rho^2 S_{FP}^2||{\mathbf{v}[n]||^4+4m\rho S_{FP} F[n]} }{4m^2g^2}})^\frac{1}{2}$ with $F[n]=||\mathbf{v}[n]||\mathbf{a}[n]\cdot\mathbf{v}[n]$. Here $m$ is the total mass of the UAV and the RIS; $\mathit{P}_{0}$, $U_{tip}^{2}$, $d_{0}$, $\rho$, $A$, and $\mathit {P_{i}}$, $S_{FP}$ are all mechanical coefficients; see \cite{2202.08486} for more details.
	
	Generally speaking, the energy consumption of the U-RIS is much larger than that of the users and the server, but the latter two are crucial in real MEC networks. Hence, we introduce a weight factor $\alpha$ for the sum of $E_{\text{p}}[n]$, with the weighted total energy consumption formulated as
	\begin{equation}\label{energy.4}
		E_{\text{total}}=\alpha\sum_{n=1}^N E_{\text{p}}[n]+\sum_{k=1}^K (	E_{k}^{\text{u}}+E_k^{\text{s}}).
	\end{equation}
	\subsection{Problem Formulation}
	In this letter, the definition of the energy efficiency is the ratio of total computed tasks in bits to the weighted total energy consumption of the system. Our main objective is to maximize the energy efficiency of the MEC system by jointly optimizing the user scheduling $\mathbf{C}\triangleq \{c_k[n], k\in\mathcal{K},n\in\mathcal{N}\}$, the phase-shift matrix of the U-RIS $\boldsymbol{\Phi}\triangleq \{\boldsymbol{\Theta}[n], n\in\mathcal{N}\}$, the U-RIS's trajectory $\mathbf{Q}\triangleq \{\mathbf{q}[n], n\in\mathcal{N}\}$, the overall computed tasks in bits $\mathbf{l}\triangleq \{{l^{\text{o}}_{k}},{l^{\text{l}}_{k}}, k\in\mathcal{K}\}$ and the CPU frequency allocation for different users of server $\mathbf{f}\triangleq \{f^{\text{o}}_k, k\in\mathcal{K}\}$. 
	The problem can be formulated as 
	\begin{subequations}\label{opt.1}
		\begin{align}
			&\underset{\mathbf{Q},\mathbf{C},\boldsymbol{\Phi},\mathbf{l},\mathbf{f}}{\max}\quad\frac{\sum_{k=1}^K ({l^{\text{o}}_{k}}+{l^{\text{l}}_{k}})}{\alpha\sum_{n=1}^N E_{\text{p}}[n]+\sum_{k=1}^K (	E_{k}^{\text{u}}+E_k^{\text{s}})}\label{optimize.1.a}\\
			&\quad~~ \textrm{s.t.} \quad~\;\, 0 \leq \theta_i[n] < 2\pi, \forall n,i, \label{optimal.1.b}\\
			&~~\quad\quad~\;\,~~~I_k\leq {l^{\text{o}}_{k}},\forall k,\label{offload.1}\\
			&~~\quad\quad~\;\,~~~\begin{matrix} \sum_{k=1}^K  f^{\text{o}}_k \end{matrix} \leq C_{\text{o}},\label{fre.1}\\
			&~~\quad \quad~\;\,~~~ \eqref{mobility.1.a}\,\rule[2.5pt]{0.1cm}{0.06em}\,\eqref{mobility.1.e},\eqref{sel.2},\eqref{toler.1}. \label{optimal.1.d}
		\end{align}
	\end{subequations}
	where \eqref{optimal.1.b} denotes the phase constraint of the U-RIS, \eqref{offload.1} indicates that there is a threshold $I_k$ for the minimum  offloading tasks for user $k$, and \eqref{fre.1} means that the total server's CPU frequency is limited by the maximum value $C_{\text{o}}$. Problem \eqref{opt.1} is a challenging mixed-integer non-linear fractional programming problem where the objective function and the constraints \eqref{sel.2} and \eqref{toler.1} are not jointly convex w.r.t. the optimization variables. In the next section, we propose an iterative algorithm to efficiently obtain a suboptimal solution.
	
	\section{Proposed Algorithm}
	In this section, we propose an iterative algorithm based on the SCA to obtain a  solution to   problem \eqref{opt.1}. Specifically, we divide (10) into two subproblems, i.e., the joint optimization of $\mathbf{C}$ and $\boldsymbol{\Phi}$ as well as that of $\mathbf{Q}$, $\mathbf{l}$ and $\mathbf{f}$. We solve the two subproblems iteratively until convergence.
	
	\subsection{Optimizing $\mathbf{C}$ and $\boldsymbol{\Phi}$ for Given $\mathbf{Q}$, $\mathbf{l}$ and $\mathbf{f}$}
	
	To handle the coupled variables, we first consider the design of the phase shift $\boldsymbol{\Phi}$. If the U-RIS chooses to serve user $k$ in time slot $n$, i.e., $c_k[n]=1$, the achievable rate is denoted as
	\begin{equation}\label{rate.2}
		R_k[n]=B\!\log_2\!\bigg(\!1\!+\!\frac{P_k\left|(\tau[n])^2\sum_{i=1}^{M}\!e^{j(\theta_i[n]+\psi_i[n])}\right|^2}{\sigma^2}\!\bigg),
	\end{equation}
	where $\psi_i[n]\!\!\!\!=\!\!\!\!\!2(m_x\!\!\!\!-\!\!\!\!1)\pi d/{\lambda}(\cos\phi_{s}[n]\sin\varphi_{s}[n]\!\!\!-\!\!\!\cos\phi_{k}[n]\sin\varphi_{k}[n])\!\!\!+\!\!\!2(m_y\!\!\!\!-\!\!\!1)\pi d/\lambda(\sin\phi_{s}[n]\sin\varphi_{s}[n]\!\!\!-\!\!\!\sin\phi_{k}[n]\sin\varphi_{k}[n])$. Clearly, if the multipath signals superimpose coherently, the rate can reach the maximum. This means that the U-RIS can play an phase alignment role  by setting $\theta_i[n]=-\psi_i[n]+{\omega},\omega \in [0,2\pi],i \in \mathcal{M}, n \in \mathcal{N}$, yielding the following maximum achievable rate
	\begin{equation}\label{rate.3}
		\setlength{\abovedisplayskip}{2pt}
		\setlength{\belowdisplayskip}{2pt}
		R_k[n]=c_k[n]B\log_2\Big(1+\frac{\xi_k}{d_{\text{s}}^{\alpha_L}[n]d_k^{\alpha_L}[n]}\Big),
	\end{equation}
	where ${\xi_k}=\frac{P_k\beta_{0}^2 M^2}{\sigma^2}$. 
	After determining the value of $\boldsymbol{\Phi}$, we relax the integer constraint \eqref{sel.2} into a linear form, and obtain the following standard LP problem for the optimization of $\mathbf{C}$: 
	\begin{subequations}\label{opt.2}
		\setlength{\abovedisplayskip}{-4.5pt}
		\begin{align}
			&\underset{{\bf C}}{ \max}\quad \sum_{n=1}^{N}\sum_{k=1}^{K} c_k[n]\check{R}_k[n]\\[-0.8mm]
			&~\textrm{s.t.} \quad~ \begin{matrix}\sum_{n=1}^N \end{matrix} c_k[n]\check{R}_k[n]   \geq \frac{N{f^{\text{o}}_k} {l^{\text{o}}_{k}}}{({f^{\text{o}}_k}T_k-{{l^{\text{o}}_{k}}\chi_k})},\forall k,\\[-0.5mm]
			&\quad\quad ~~\,\begin{matrix}\sum_{k=1}^K\end{matrix} c_k[n]=1,\,0 \leq c_k[n]\leq 1,\forall k,n,
		\end{align}
	\end{subequations}
	where $\!\check{R}_k[n]\!\!=\!\!B\log_2(1\!+\!\frac{\xi_k}{d_{\text{s}}^{\alpha_L}[n]d_k^{\alpha_L}[n]})$. This problem can be solved by the CVX solver. Finally, $c_k[n],\forall k,n,$ are reconstructed as binary variables via rounding.
	
	\subsection{Optimizing $\mathbf{Q}$, $\mathbf{l}$ and $\mathbf{f}$ for Given $\mathbf{C}$ and $\boldsymbol{\Phi}$}
	For any given $\mathbf{C}$ and $\boldsymbol{\Phi}$, problem \eqref{opt.1} can be recast as
	\begin{subequations}\label{opt.3}
			\setlength{\abovedisplayskip}{3pt}
		\begin{align}
			&\underset{\mathbf{Q},\mathbf{l},\mathbf{f}}{\max}\quad\frac{\sum_{k=1}^K ({l^{\text{o}}_{k}}+{l^{\text{l}}_{k}})}{\alpha\sum_{n=1}^N E_{\text{p}}[n]+\sum_{k=1}^K (	E_{k}^{\text{u}}+E_k^{\text{s}})}\label{optimize.3.a}\\
			&~\textrm{s.t.} \quad~ \eqref{mobility.1.a}\,\rule[2.5pt]{0.1cm}{0.06em}\,\eqref{mobility.1.e},\eqref{toler.1},\eqref{offload.1},\eqref{fre.1}
		\end{align}
	\end{subequations}
	
	Note that problem \eqref{opt.3} is difficult to solve due to the non-convex objective function and constraint \eqref{toler.1}. To obtain an approximate solution, we first consider convexifying the constraint. By introducing the slack variables ${\mathbf y}=\left\{y_k[n],\forall k,n\right\}$, $ {\bf p}=\left\{p[n],\forall n\right\}$, where $y_k[n] \geq \parallel {\bf q}[n]-{\bf {w}}_k\parallel^2+H^2$, $p[n] \geq \parallel{\bf q}[n]-{\bf {w}}_{\text{s}}\parallel^2+H^2$, and the auxiliary variable ${\bf D}=\{d_k,\forall k\}$,  constraint \eqref{toler.1} can be transformed into
	\begin{subequations}
		\setlength{\abovedisplayskip}{2pt}
		\begin{align}
			&d_k \leq\frac{1}{N}\sum_{n=1}^N c_k[n]B\gamma_0[n],\forall k,\label{cons.1}\\[-1mm]
			&\frac{{l^{\text{o}}_{k}}^2}{d_k}+\chi_k\frac{{l^{\text{o}}_{k}}^2}{f^{\text{o}}_k} 	\leq T_k {l^{\text{o}}_{k}},\forall k,\label{cons.2}\\[-0.5mm]
			&{l^{\text{l}}_{k}}\chi_k \leq T_k f^{\text{l}}_k,\label{cons.3}\forall k, 
		\end{align}
	\end{subequations}
	where $\gamma_0[n]=\log_2(1+\frac{\xi_k}{(p[n])^{\alpha_L/2}(y_k[n])^{\alpha_L/2}})$. Since $y_k[n]$ and $p[n]$ can be increased to reduce the objective value, the constraints of $y_k[n]$ and $p[n]$ must hold with equality at the optimal solution to problem \eqref{opt.3}; hence, constraint (5) can be equivalently relaxed into (15) without loss of optimality. Note that $\gamma_0[n]$ is convex with respect to $y_k[n]$ and $p[n]$, so we apply the first-order Taylor expansion of $\gamma_0[n]$ at the given point $(p^{(t)}[n]$, $y_k^{(t)}[n])$ in the $t$th iteration to convert the non-convex constraint \eqref{cons.1} to a convex form as follows
	\begin{align}\label{cons.4}
		d_k \leq \frac{1}{N}\sum_{n=1}^N c_k[n]B{\hat{R}_k[n]},\forall k,
	\end{align}
	where
	\begin{align*}
		{\hat{R}_k}\,[n]\,&=C_k^{(t)}[n]\!+\! A_k^{(t)}[n](p[n]-p^{(t)}\![n])\!+\!B_k^{(t)}[n]\mathit{Y}_k^{(t)}[n]\notag \\
		A_k^{(t)}[n]&=-\log_{2}e\frac{({\alpha_L}{/}{2})\xi_k}{{p^{(t)}[n]}^{(\alpha_L/2+1)}y_k^{(t)}{[n]}^{(\alpha_L/2)}+\xi_k p^{(t)}[n]}\notag \\
		B_k^{(t)}[n]&=-\log_2 e\frac{(\alpha_L/2)\xi_k}{p^{(t)}{[n]}^{(\alpha_L/2)}y_k^{(t)}{[n]}^{(\alpha_L/2+1)}+\xi_k y_k^{(t)}[n]}\notag \\
		C_k^{(t)}[n]&=\log_2(1+\frac{\xi_k}{p^{(t)}{[n]}^{(\alpha_L/2)}y_k^{(t)}{[n]}^{(\alpha_L/2)}})\notag 
		\end{align*}
	and $\mathit{Y}_k^{(t)}[n]=(y_k[n]-y_k^{(t)}[n])$.
	
	We next deal with the non-convex objective function. Note that $E_{\text{p}}[n]$ and $E_k^{\text{s}}$ are the two non-convex terms in the denominator of the objective function. To tackle the non-convexity of $E_{\text{p}}[n]$, we successively apply the inequalities $ \boldsymbol{a}\!\cdot\!\mathbf{b}\!\leq\!||\boldsymbol{a}||||\mathbf{b}||,\boldsymbol{a},\mathbf{b}\!\in\!{{\mathbb{R}^2}}$ and $(a^2+b^2)^\frac{1}{2}\!\leq\! (a+b),a,b\in {{\mathbb{R}_+}}$ for the non-convex term $F[n]$ and $\sqrt{((\kappa[n])^2+\frac{||\mathbf{v}[n]||^4}{4v_{0}^{4} } )^{\frac{1}{2} } -\frac{||\mathbf{v}[n]||^2}{2v_{0}^{2}}}$ in $E_{\text{p}}[n]$, respectively. Finally we get a convex upper bound $E_{\text{up}}[n]$ of $E_{\text{p}}[n]$, expressed as
	\begin{align}\label{Ep.1}
		E_{\text{up}}[n] 
		\! \!=\!\! P_{0}\Big(\!1\!\!+\!\!\frac{3||\mathbf{v}[n]||^2}{U_{tip}^{2} } \!\Big)\!\!+\!\!\frac{1}{2}d_{0}\rho sA||\mathbf{v}[n]||^3\!\!+\!\! P_{i}(\hat{\kappa}[n])^2\!.  
	\end{align}
To handle the non-convexity of $E_k^{\text{s}}$, similarly we introduce the slack variable $\mathbf{u}=\{u_k,\forall k\}$ satisfying  $u_k \geq {l^{\text{o}}_{k}}(f^{\text{o}}_k)^{2}$. By   inequality transformation and applying first-order Taylor expansion at given point ${l^{\text{o}}_{k}}^{(t)}$ in the $t$th iteration, we have
	\begin{equation}\label{Es.1}
		\frac{1}{{l^{\text{o}}_{k}}^{(t)}}+({l^{\text{o}}_{k}}-{{l^{\text{o}}_{k}}^{(t)}})\Big(-\frac{1}{({{l^{\text{o}}_{k}}^{(t)}})^2}\Big) \geq \frac{(f^{\text{o}}_k)^2}{u_k},\forall k.
	\end{equation}
	Therefore, problem \eqref{opt.3} can be approximated as 
	\begin{subequations}\label{opt.4}
		\begin{align}
			&\underset{\begin{scriptsize}\begin{array}{c}
						\mathbf{Q},\!\mathbf{y},\!\mathbf{p}\\
						\!\mathbf{u},\!\mathbf{l},\!\mathbf{f}\\
			\end{array}\end{scriptsize}}{\max}~\, \frac{\sum_{k=1}^K {l^{\text{o}}_{k}}+{l^{\text{l}}_{k}}}{\alpha \sum_{n=1}^N E_{\text{up}}[n]+\sum_{k=1}^K(	E_{k}^{\text{u}}+{\varphi_{\text{s}}}\chi_k u_k)} \label{opt.4.1}\\
			&~~~\,\textrm{s.t.} \quad~ y_k[n] \geq \parallel {\bf q}[n]-{\bf w}_k\parallel^2+H^2,\forall k,n, \label{opt.4.2} \\
			&~~\quad\quad\,~\;\;p[n] \geq \parallel{\bf q}[n]-{\bf w}_{\text s}\parallel^2+H^2,\forall n,\label{opt.4.3}\\
			&~~\quad\quad\,~\;\,\eqref{mobility.1.a}\,\rule[2.5pt]{0.1cm}{0.06em}\,\eqref{mobility.1.e},\!\eqref{offload.1},\!\eqref{fre.1},\!\eqref{cons.2},\!\eqref{cons.3},\!\eqref{cons.4},\!\eqref{Es.1}. \label{opt.4.4}
		\end{align}
	\end{subequations}
	This problem is quasi-convex because the objective function consists of a linear numerator as well as a convex denominator and all constraints are convex. Therefore, it can be efficiently solved by existing fractional programming methods, such as the Dinkelbach algorithm.

	\subsection{Overall Algorithm}
	Based on the solutions obtained in the previous two subsections, the designed overall algorithm for problem \eqref{opt.1} is summarized in Algorithm \ref{alg1}. The complexity of the proposed algorithm is $\mathcal{O}(N^{3.5}\log({1}/{\epsilon}))$.
	In general, Algorithm \ref{alg1} yields a lower bound of the original problem, and its performance in improving system energy efficiency is verified in Section IV.
	
	\begin{algorithm}[t] 
		\caption{Proposed algorithm for solving problem \eqref{opt.1}} 
		\label{alg1} 
		\begin{algorithmic}[1] 
			\STATE \textbf{Initialization}: Initial $\{\mathbf{Q}_0,\mathbf{l}_0,\mathbf{f}_0,\mathbf {y}_0,\bf {p}_0,\mathbf{u}_0\}$ and iteration number $t=0$.
			\STATE \textbf{repate}
			\STATE \quad Update $\mathbf{C}_{t+1}$ with given $\{\mathbf{Q}_t,\mathbf{l}_t,\mathbf{f}_t\}$ by solving \eqref{opt.2};
			\STATE\quad Update$\{\mathbf{Q}_{t+1},\mathbf{l}_{t+1},\mathbf{f}_{t+1},\mathbf {y}_{t+1},\bf {p}_{t+1},\mathbf{u}_{t+1}\}$ with given   $\quad\quad~~\;$ $\{\mathbf{C}_{t+1},\mathbf{Q}_t,\mathbf{l}_t,\mathbf {y}_{t},\bf {p}_{t},\mathbf{u}_{t}\}$ by solving  \eqref{opt.4};
			\STATE \quad Update $t=t+1$;
			\STATE \textbf{until}: The value of objective function for problem \eqref{opt.1} converges to a predetermined accuracy $ \epsilon$.
			\STATE \quad Determine $\boldsymbol{\Phi}$ with finalized $\mathbf{Q}$ and $\mathbf{C}$ by the phase $\quad\quad~~\;$ alignment method.
		\end{algorithmic} 
	\end{algorithm}
	
	\section{Numerical Results}
	\begin{figure*}[t]
		\setlength{\belowcaptionskip}{-20pt}
		\setlength{\abovecaptionskip}{-0.1cm} 
		\centering
		\begin{minipage}{0.325\linewidth}
			\centering
			\includegraphics[width=6.6cm, height=2.0in]{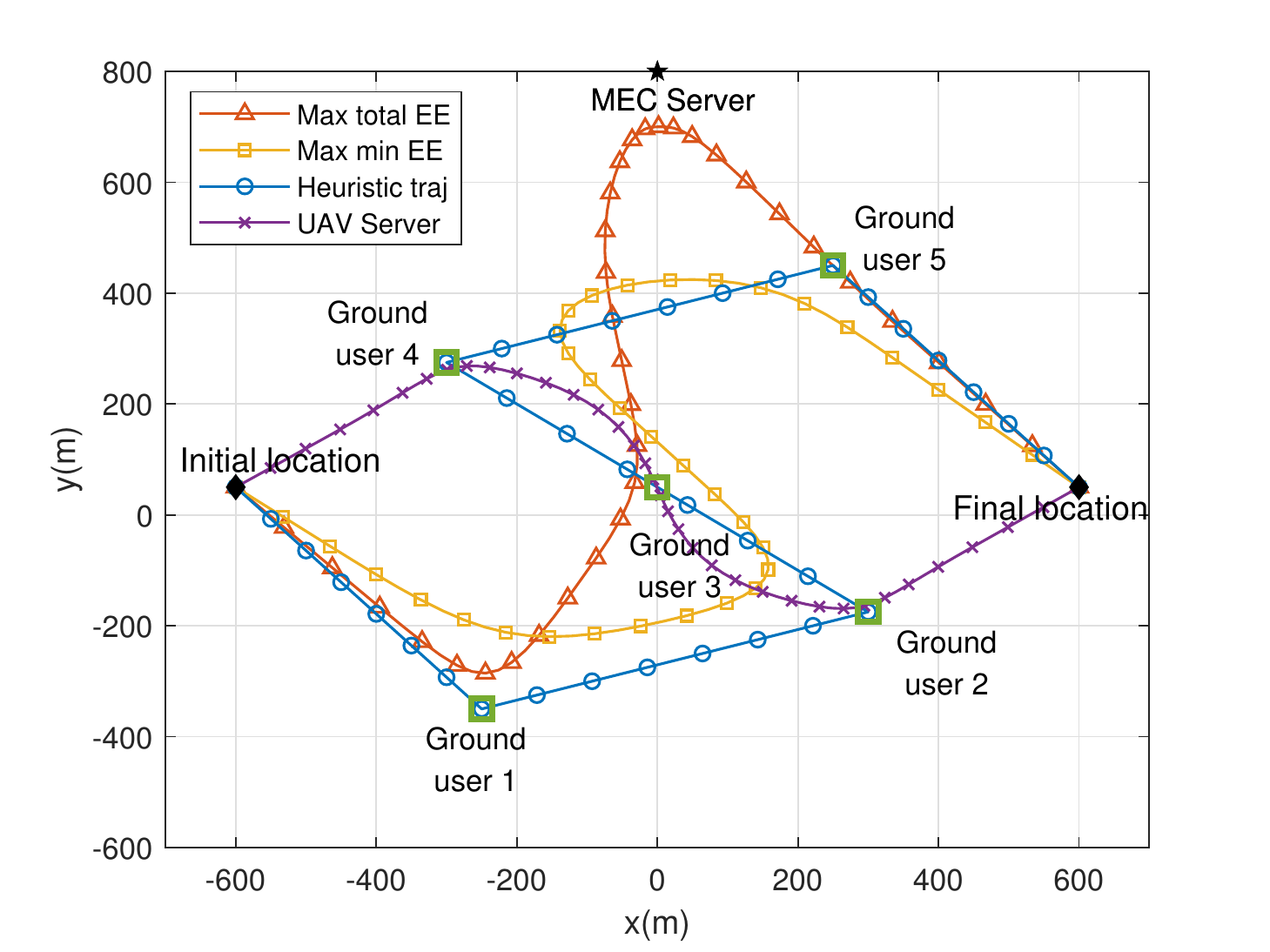}
			\caption{Trajectories of algorithms.}
			\label{fig1}
		\end{minipage}
		\begin{minipage}{0.325\linewidth}
			\centering
			\includegraphics[width=6.6cm, height=2.0in]{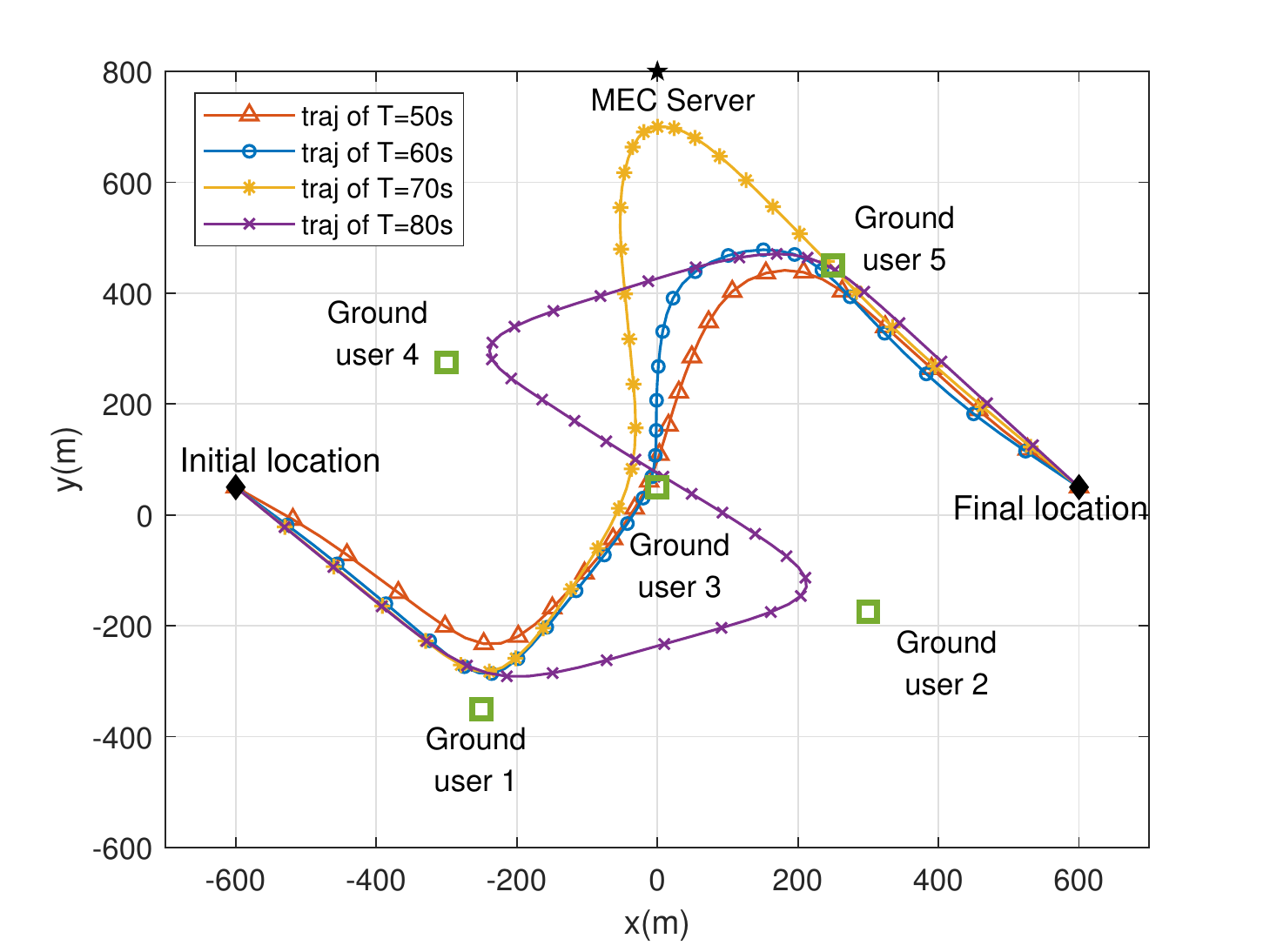}
			\caption{Trajectories of different $T$.}
			\label{fig2}
		\end{minipage}
		\begin{minipage}{0.325\linewidth}
			\centering
			\includegraphics[width=6.6cm, height=2.0in]{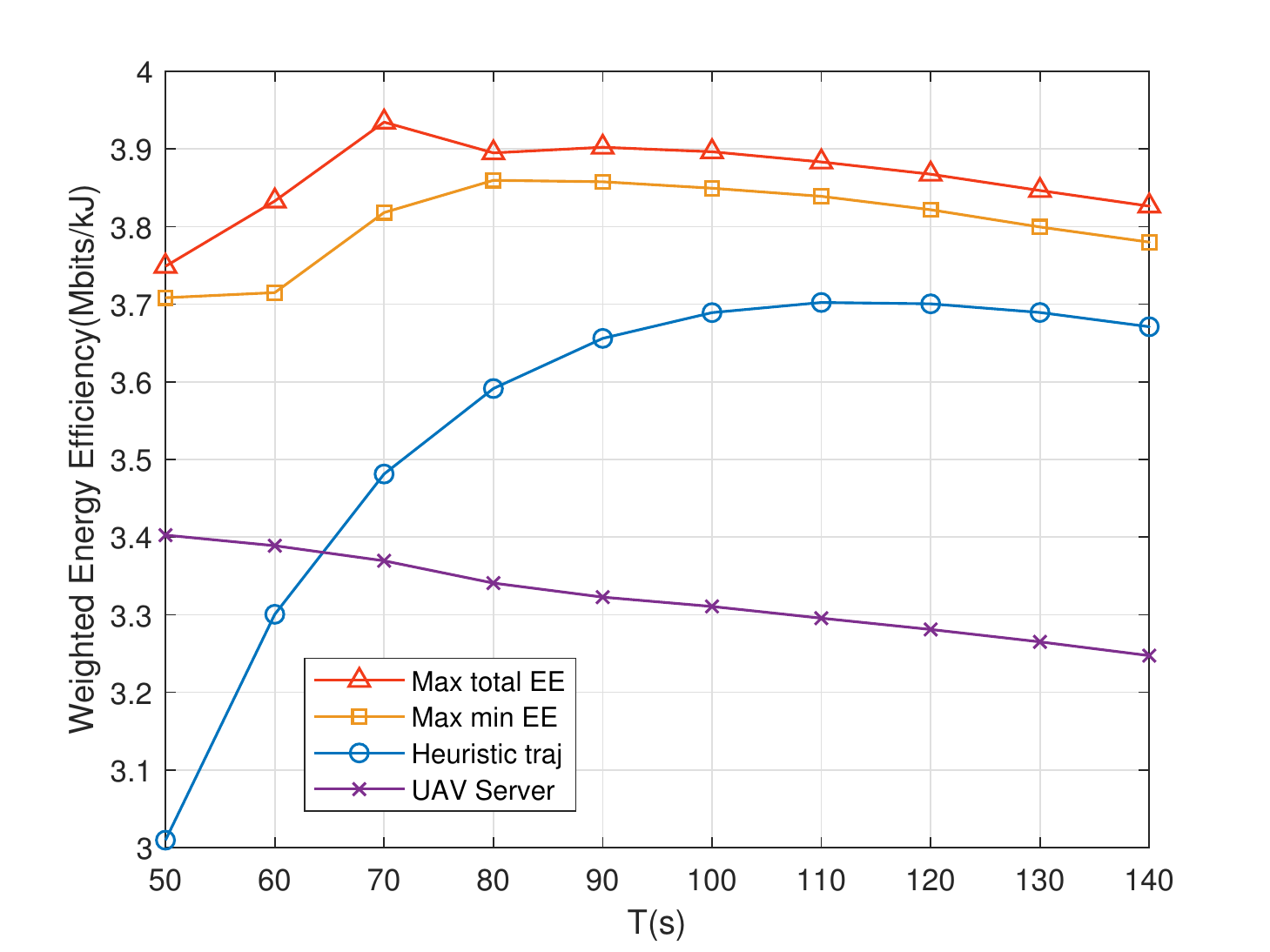}
			\caption{Weighted EE versus $T$.}
			\label{fig3}
		\end{minipage}
		\vspace{-6mm}
	\end{figure*}
	In this section, we analyze the effectiveness of the proposed algorithm with numerical results. The objective of the proposed algorithm is to maximize the total energy efficiency of the whole MEC system (denoted as max total EE), at the expense of the quality of task processing for some individual users. When each user has urgent tasks and it is important to ensure high energy efficiency and fairness, we also propose an algorithm to maximize the minimum energy efficiency over all users (denoted as max min EE) by replacing the objective function in \eqref{opt.1} with$\frac{l_{\min}}{\alpha\sum_{n=1}^N E_{\text{p}}[n]+\sum_{k=1}^K (	E_{k}^{\text{u}}+E_k^{\text{s}})}$ where $l_{\min} \leq {l^{\text{o}}_{k}}+{l^{\text{l}}_{k}},\forall k$. This problem can be solved similarly, and the algorithm will not be described due to space limitations. 
	
	Besides, we set two baseline algorithms aiming to maximize the system total energy efficiency for comparison: 1) Heuristic-traj means that the U-RIS traverses each user node along the pre-defined shortest route at a constant speed, with optimized $\mathbf{C},\boldsymbol{\Phi},\mathbf{l}$ and $\mathbf{f}$; 2) UAV-Server means that
	 a UAV-carried server is used to provide edge computing, with optimized $\mathbf{C},\mathbf{Q},\mathbf{l}$ and $\mathbf{f}$. This is the traditional practice of the UAV-assisted MEC system. Compared with the proposed algorithm, UAV-Server will not lead to a cascade channel but increase the flight burden. Besides, the maximal CPU frequency of the MEC server in the UAV-Server scheme is lower than that of the land-based MEC server in practice.
	Simulation parameters are set as $B=$ 1 MHz, $\sigma^2=-160$ dBm,  $\beta_{0}=-40$ dB, $M=10^3$, $H=100$ m, ${v}_{max}=50$ m/s, ${a}_{max}=30 ~\text{m}/{\text{s}^2}$, ${\delta_t}=1$ s, ${\bf q}_0=[-600,50]^T$ m, ${\bf q}_F=[600,50]^T$ m, ${\bf w}_{\text s}=[0,800]^T$ m, $r_d=700$ m, ${\varphi_{\text{u}}}=10^{-8}$, ${\varphi_{\text{s}}}=10^{-5}$, $C_{\text{o}}=3000$ MHz, $\alpha=0.02$. $P_k=0.1 ~\text{W}$, $\chi_k=10^3~\text{cycles/bit}$, $f^{\text{l}}_k=100~\text{MHz}$, $T_k=30~\text{s}$, $I_k=1~\text{Mbits}$, for $\forall k$,
	the mass of UAV, RIS and MEC server are  2 kg, 2 kg and 20 kg, respectively.
	
	Fig.~\ref{fig1} shows the U-RIS or UAV trajectories of the various algorithms ($T=70$\,s), all of which are significantly different. We observe that in the proposed max total EE algorithm, the U-RIS first flies towards user 1, and then approaches the server along the trajectory colse to user 3. After that, the U-RIS decelerates and finally reaches the endpoint. Flying along this trajectory achieves the highest total energy efficiency because it strikes the best balance between the G-U and U-S links, i.e., an effective trade-off is made between ``near the server to enjoy high-quality channel with all user'', and ``poor quality channel during flying towards the remote server''. In this algorithm, the U-RIS only meets the minimum demand $I_k$ of the poor users, while serving the users with high channel quality to increase the total processed tasks. On the contrary, the U-RIS in max min EE needs to take account of each user, especially users 1 and 2, whose long-distance from the server leads to greater path loss. For the Heuristic-traj algorithm, the U-RIS flies along a straight-line route through all users, which consumes a lot of energy due to the frequent speed and direction changes. Besides, as for the UAV-Server algorithm, compromises are made between the flight energy consumption and the channel quality. The UAV in this scheme flies to ${\bf q}_F$  focusing on the service for users 2, 3 and 4 with fewer detours.
	
	Fig.~\ref{fig2} and Fig.~\ref{fig3} illustrate the trajectories of the max total EE algorithm and the  total EE of various algorithms versus $T$, respectively. Different from the characteristics of a general UAV-assisted system, users in the MEC system have their own tolerated latency. Hence the excessive flight time $T$ will rather lead to a decrease of the EE (as shown in Fig.~\ref{fig3} when $T \geq 110$\,s), so we set a moderate period $T=50-80$\,s to perform the trajectory analysis. In Fig.~\ref{fig2}, the trajectory of $T=70$\,s has a visible difference from that of other $T$, because it has just achieved the trade-off mentioned above, which is also the reason why the EE peaks at 70\,s in the EE curve of max total EE algorithm in Fig.~\ref{fig3}. This means that 70\,s is the optimal time span of this system. As the period grows longer ($T=80$\,s), it is necessary to make full use of the latency of each user to offload more tasks. Hence the U-RIS adopts a strategy similar to max min EE, and the EE of these two algorithms get close when $T\geq 80$\,s. In Fig.~\ref{fig3}, it is observed that the proposed max total EE algorithm has a considerable performance gain compared to the baseline algorithms. It is worth noting that the Heuristic-traj algorithm can be treated as a lower bound of the EE for the preliminary system design. Besides, due to the high flight energy consumption of the UAV-Server scheme, the EE under this algorithm shows a downward trend even in the early stage.
		
	\section{Conclusions}
	In this paper, the UAV-mounted RIS technique has been introduced to enhance the user offloading link, thereby further improving the performance of the MEC system. We proposed a joint RIS passive beamforming, UAV trajectory and MEC resource allocation optimization algorithm to maximize the EE and obtained a high-quality suboptimal solution. Numerical results gave evidence of the significant benefits that the U-RIS can provide in the MEC system.

	\bibliographystyle{IEEEtran}
	\bibliography{Lireference}

\end{document}